\documentclass{jpsj3}
\usepackage{color}

\title{
Direct observation of the ground state of 
a 1/3 quantum magnetization plateau 
in SrMn$_3$P$_4$O$_{14}$
using neutron diffraction measurements
}

\author{
Masashi Hase$^1$\thanks{HASE.Masashi@nims.go.jp}, 
Vladimir Yu. Pomjakushin$^2$, 
Andreas D\"onni$^1$, 
Tao Yang$^3$, 
Rihong Cong$^3$, and 
Jianhua Lin$^3$
}

\inst{
$^1$National Institute for Materials Science (NIMS), 
Tsukuba, Ibaraki 305-0047, Japan \\
$^2$Laboratory for Neutron Scattering, Paul Scherrer Institut (PSI), 
CH-5232 Villigen PSI, Switzerland \\
$^3$College of Chemistry and Molecular Engineering, 
Peking University, Beijing 100871, People's Republic of China
} 

\abst{
We can directly investigate  
the ground state in 
magnetization-plateau fields (plateau ground state)
using neutron diffraction measurements.
We performed neutron diffraction measurements 
on the spin-5/2 trimer substance 
SrMn$_3$P$_4$O$_{14}$ 
in magnetization-plateau fields.
The integrated intensities of magnetic reflections 
calculated using an expectation value of each spin in a plateau ground state 
of an isolated-trimer model 
agree well with
those obtained experimentally 
in the magnetization-plateau fields.  
We succeeded in direct observation of a plateau ground state  
in SrMn$_3$P$_4$O$_{14}$. 
}

\begin{document}
\maketitle

\section{Introduction}

During typical quantum-mechanics calculations, 
we first calculate eigenstates and eigenenergies of a model Hamiltonian and 
then calculate physical quantities such as magnetization and specific heat. 
We compare the experimentally obtained physical quantities   
of a substance and 
those calculated in a hypothesized model Hamiltonian. 
Then we judge whether the hypothesized model is applicable to the substance. 
We do not usually investigate the eigenstates themselves 
directly in experiments.

We can directly investigate 
a ground state in magnetic fields 
where a quantum magnetization plateau appears 
(plateau ground state)
using neutron diffraction measurements 
as explained below.\cite{Comment1} 
Before providing an explanation, 
we describe a quantum magnetization plateau.\cite{Hida94,Oshikawa97}
The magnetization of a plateau ground state 
takes its maximum value because of the magnetic fields. 
Excited states cannot contribute to magnetization because of 
energy gaps separating the plateau ground state and excited states. 
Therefore, magnetization cannot increase and 
the quantum magnetization plateau appears. 

The existence of an energy gap is important for 
the occurrence of a quantum magnetization plateau. 
It can be confirmed using other methods. 
A zero-magnetization plateau appears in spin systems possessing 
spin-singlet ground state(s) and 
a spin gap separating the ground and first-excited states. 
Examples are Haldane substances,\cite{Katsumata89,Ajiro89} 
the spin-Peierls cuprate CuGeO$_3$,\cite{Hase93c} and 
the orthogonal-dimer compound SrCu$_2$(BO$_3$)$_2$.\cite{Kageyama99} 
It is possible to prove 
the existence of a spin gap explicitly by 
the exponential decay of magnetic susceptibility 
on cooling.\cite{Kageyama99,Renard87,Hase93a,Hase93b} 
Inelastic neutron scattering (INS) techniques are useful 
not only for a spin gap generating a zero-magnetization plateau, 
but also for an energy gap generating a finite-magnetization plateau.
The existence of an energy gap generating a finite-magnetization plateau 
was confirmed in NH$_4$CuCl$_3$\cite{Ruegg04}, 
Cu$_3$(P$_2$O$_6$OD)$_2$\cite{Hase07}, and 
SrMn$_3$P$_4$O$_{14}$.\cite{Hase11b}

In a plateau ground state, 
a magnetic moment on each site must be fully polarized 
parallel and anti-parallel to external magnetic fields 
when $\langle S_{jz} \rangle$ value on the site is positive and negative, 
respectively.\cite{Comment2}
Here, $\langle S_{jz} \rangle$ is 
an expectation value of spin on an $j$-th magnetic-ion site 
in the plateau ground state. 
The full polarizations generate the magnetization plateau.
We can determine a value of $g \mu_{\rm B} \langle S_{jz} \rangle$  
(a value of a magnetic moment)  
in magnetization-plateau fields from magnetic reflections  
obtained in neutron diffraction measurements. 
Therefore, we can compare theoretical and experimental values of 
$\langle S_{jz} \rangle$, 
meaning that direct investigation of a plateau ground state 
can be done through 
neutron diffraction measurements.

Magnetization plateaus, however, are usually apparent 
in high magnetic fields or at extremely low temperatures. 
Therefore, 
neutron diffraction measurements of plateau ground states 
are usually difficult. 
They have not been reported in the literature to date. 

A 1/3 quantum magnetization plateau is apparent  
between 2 and 10 T at 1.3 K in 
insulating SrMn$_3$P$_4$O$_{14}$.\cite{Yang08,Hase09a}
Therefore, it is possible to perform neutron diffraction measurements of 
the substance in magnetization-plateau fields. 
Here, we describe the crystal structure and 
magnetism of SrMn$_3$P$_4$O$_{14}$. 
The space group is monoclinic $P2_1 /c$ (No. 14).\cite{Yang08}  
The lattice constants at 4.0 K are
$a = 7.661(1)$ \AA, $b = 7.784(1)$ \AA, $c = 9.638(1)$ \AA , 
and $\beta = 111.70(2)^{\circ}$.\cite{Hase11a} 
Two Mn$^{2+}$ sites (Mn1 and Mn2) exist as presented in Fig. 1. 
The electron configuration in Mn$^{2+}$ ions is $3d^5$.
Anisotropy in exchange interactions is caused by 
the spin-orbit interaction and a low symmetry of crystal fields. 
The spin value on Mn$^{2+}$ ions in this substance is 5/2.\cite{Hase09a} 
Therefore, the orbital moment is 0. 
The spin-orbit interaction and single ion anisotropy 
do not exist.  
The $g$-value is isotropic and about 2. 
We evaluated experimentally that 
the powder-average $g$-value was 1.98 
from the saturated value of magnetization ($4.95 \mu_{\rm B}$ per Mn). 
The Mn atoms are coordinated octahedrally by six oxygen atoms. 
The symmetry of crystal fields affecting the Mn$^{2+}$ ions 
is nearly cubic.
Accordingly, the anisotropy in exchange interactions of 
Mn spins in this substance is expected to be small. 
We can consider that 
spin-5/2 on Mn$^{2+}$ ions is a Heisenberg spin. 

\begin{figure}
\begin{center}
\includegraphics[width=8cm]{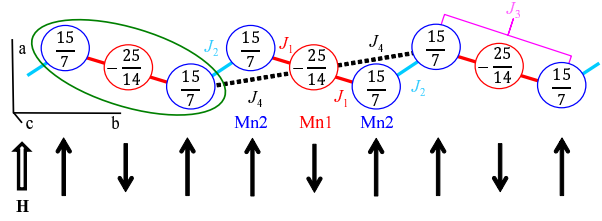}
\caption{
(Color online)
The spin system in SrMn$_3$P$_4$O$_{14}$.\cite{Hase09a} 
Mn$^{2+}$ ions ($3d^5$) have localized spin 5/2. 
Two kinds of Mn sites (Mn1 and Mn2) exist. 
The major AF $J_1$ interaction forms spin trimers 
as shown by the ellipse. 
Considering $J_i$ ($i = 1 \sim 4$) interactions, 
we can explain a coplanar spiral magnetic structure  
below $T_{\rm N} = 2.2(1)$ K in the zero magnetic field.\cite{Hase11a} 
The values on Mn sites show $\langle S_{jz} \rangle$ 
in the plateau ground state in the isolated trimer.
Arrows depict schematically  
magnetic moments ($g \mu_{\rm B} \langle S_{jz} \rangle$)  
in the plateau ground state obtained in the present study.  
}
\end{center}
\end{figure}

We can explain magnetizations\cite{Hase09a} and 
magnetic excitations observed in INS experiments\cite{Hase11b} 
using an isolated antiferromagnetic (AF) Heisenberg spin-5/2 trimer. 
The trimer indicated by the ellipse in Fig. 1
is formed by the AF $J_1$ interaction. 
The value of $J_1$ was estimated as 
$3.4 \sim 4.0$ K.\cite{Hase09a,Hase11b}
A coplanar spiral magnetic structure appears 
below $T_{\rm N} = 2.2(1)$ K in the zero magnetic field.\cite{Hase11a} 
Therefore, other exchange interactions are not negligible. 
The spiral magnetic structure results from  
frustration between nearest-neighbor and 
next-nearest-neighbor exchange interactions 
in the trimerized chain formed by $J_i$ ($i = 1 \sim 4$) interactions. 

In the isolated-trimer model, 
states with 
$ \langle ({\bf S}^{\rm T})^2 \rangle = \frac{5}{2} ( \frac{5}{2} + 1)$ 
and $E/J_1 = -15$ are 
six-folded ground states in 0 T  
where ${\bf S}^{\rm T}$ and $E$ represent 
the sum of three spin operators in a trimer and 
eigenenergy per three spins, respectively. 
In magnetic fields up to 10 T, 
one ground state with $ \langle S^{\rm T}_z \rangle = 5/2$ 
is the unique ground state.  
In the ground state, 
each $\langle S_{jz} \rangle$ value was calculated as 
$\langle S_{1z} \rangle = -25/14$ and 
$\langle S_{2z} \rangle = 15/7$
on Mn1 and Mn2 sites, respectively.  
As shown in Fig. 1, 
the spin on Mn1 (Mn2) site is anti-parallel (parallel) to magnetic fields 
in magnetization-plateau fields. 
Therefore, the 1/3 magnetization plateau appears 
($ \langle S^{\rm T}_z \rangle = 
\langle S_{1z} \rangle + 2 \langle S_{2z} \rangle = 5/2$).

We performed neutron diffraction measurements  
of SrMn$_3$P$_4$O$_{14}$ in magnetic fields. 
In analyses, we assumed that 
the plateau ground state in SrMn$_3$P$_4$O$_{14}$
was close to that in the isolated AF spin-5/2 trimer. 
Experimental integrated intensities of magnetic reflections 
are consistent with 
integrated intensities of magnetic reflections calculated 
using $\langle S_{jz} \rangle$ values 
expected in the isolated trimer 
($\langle S_{1z} \rangle = -25/14$ and 
$\langle S_{2z} \rangle = 15/7$). 

\section{Methods of Experiments}

Single crystals of SrMn$_3$P$_4$O$_{14}$ were synthesized 
under hydrothermal conditions at 473 K.
Details of the synthesis have been reported elsewhere.\cite{Yang08}  
Each crystal was small. 
For that reason, we pulverized crystals. 
We entered powders in paraffin molten by heating and 
embedded the powders in solid paraffin. 
We measured magnetizations of the powders embedded in solid paraffin
up to $H = 5$ T and down to 1.8 K 
using a superconducting quantum interference device 
(SQUID) magnetometer (MPMS-5S; Quantum Design). 

We conducted neutron diffraction experiments 
at the Swiss spallation neutron source (SINQ) 
in Paul Scherrer Institut (PSI).  
We used the high-resolution powder diffractometer for thermal neutrons HRPT.\cite{hrpt} 
The wave length $\lambda$ was 2.45 \AA. 
Magnetic fields of 0 - 6 T were applied 
almost perpendicular to the scattering plane 
using a superconducting magnet.
The deviation from $90^{\circ}$ 
of the angle between the scattering vector ${\bf Q}$ and magnetic field ${\bf H}$
is less than $6^{\circ}$ ($\cos 6^{\circ} = 0.995$). 
We used pressed pellets of SrMn$_3$P$_4$O$_{14}$ 
to minimize the problem of powder realignment in strong magnetic fields.
We performed Rietveld refinements of the crystal structure
using the {\tt FULLPROF Suite}  program package,\cite{Rodriguez93}  
using its internal tables for scattering lengths. 

\section{Results and discussion}

The circles in Fig. 2(a) show a neutron diffraction pattern of 
SrMn$_3$P$_4$O$_{14}$ pellets    
at 20 K in 0 T (paramagnetic state). 
We performed Rietveld refinements of the crystal structure. 
We assume that directions of domains are randomly distributed 
both in the present (pellet) and previous (powder) samples.\cite{Hase11a} 
The line on the experimental pattern 
indicates the result of Rietveld refinements.  
It can explain the experimental pattern. 
Structural parameters obtained in the present refinements 
are consistent with those obtained in the previous refinements. 
Therefore, no alignment of powders occurs in producing the pellets. 
Figure 2(b) shows diffraction patterns of pellets at 1.6 K in 0 and 6 T. 
The two diffraction patterns are mutually close at high angles, 
indicating that  
no alignment of powders in the pellets occurs 
by application of magnetic fields.

\begin{figure}
\begin{center}
\includegraphics[width=8cm]{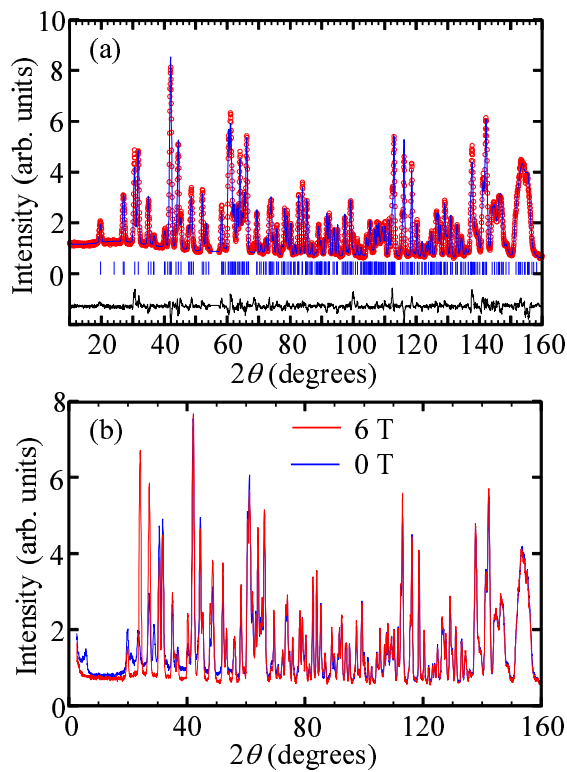}
\caption{
(Color online)
Neutron diffraction patterns of SrMn$_3$P$_4$O$_{14}$ pellets 
measured using the HRPT diffractometer ($\lambda = 2.45$ \AA).
(a) 
Circles show a diffraction pattern measured at 20 K in 0 T. 
Lines on the measured pattern and at the bottom 
portray a Rietveld refined pattern and 
the difference between the measured and the Rietveld refined patterns, respectively. 
Hash marks represent the positions of nuclear reflections.
(b)
Neutron diffraction patterns at 1.6 K in 0 and 6 T. 
}
\end{center}
\end{figure}

Figure 3 depicts fragments of neutron diffraction patterns of 
SrMn$_3$P$_4$O$_{14}$ pellets  
at 1.6 K in several magnetic fields. 
Open blue circles under the patterns denote positions of 
major magnetic reflections generated by 
the spiral magnetic structure.\cite{Hase11a} 
The intensities decrease concomitantly with increasing magnetic field $H$ and 
disappear in $H \geq 0.3$ T.
The positions of the magnetic reflections are independent of $H$.
The spiral magnetic order results from 
the AF $J_1$ interaction ($3.4 \sim 4.0$ K) and other weaker interactions 
in the quasi-one-dimensional system in Fig. 1. 
The order is easily destroyed by the weak magnetic fields. 

\begin{figure}
\begin{center}
\includegraphics[width=8cm]{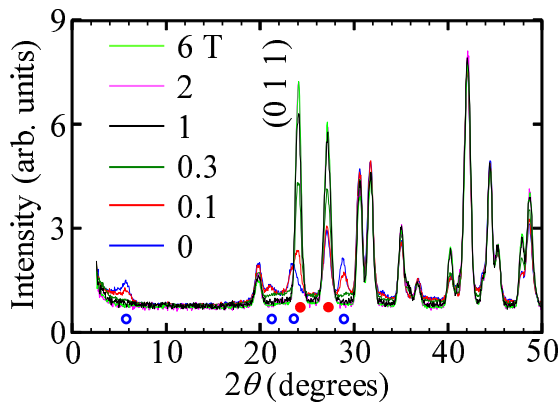}
\caption{
(Color online)
Neutron diffraction patterns of SrMn$_3$P$_4$O$_{14}$ pellets 
at 1.6 K in several magnetic fields 
measured using the HRPT diffractometer ($\lambda = 2.45$ \AA). 
Open blue circles represent positions of 
major magnetic reflections generated by the spiral magnetic structure. 
Closed red circles represent positions of 
major magnetic reflections with integer indices 
appearing in $H \geq 0.05$ T. 
}
\end{center}
\end{figure}

Closed red circles under the patterns in Fig. 3 
indicate positions of major new magnetic reflections with integer indices 
appearing in $H \geq 0.05$ T. 
Figure 4 shows the integrated intensity of the (0 1 1) reflection 
at $2 \theta = 24.1^{\circ}$.   
Figure 4(a) shows that  
the intensity is strong in high magnetic fields and at low temperatures. 
Figure 4(b) shows that
the intensity at 1.6 K (open red circle) increases concomitantly 
with increasing $H$ and that 
it is almost independent of $H$ above 2 T. 
The $H$ dependence is similar to that of square of 
magnetization measured at 1.8 K (closed blue circle). 
Therefore, the new reflections are 
correlated with the 1/3 quantum magnetization plateau. 
Figure 4(c) shows that  
the intensity in 6 T (open red circle) decreases 
with increasing temperature $T$ and that
it is negligible at 20 K. 
The $T$ dependence of a normalized order parameter 
is usually expressed as  
$(1-T/T_{\rm c})^{\beta} \ (0< \beta <1)$
just below a transition temperature $T_{\rm c}$.
The $T$ dependence 
of the intensity in 6 T
differs from that of an order parameter, 
indicating that no phase transition occurs. 
The state in $H \geq 0.3$ T is a polarized paramagnetic state 
without spontaneous magnetic order.
The magnetic reflections are generated by 
magnetic moments aligned parallel or anti-parallel 
to the applied magnetic fields.  
The $T$ dependence of the (0 1 1) intensity is inconsistent with 
that of square of magnetization measured in 5 T (closed blue circle). 
Magnetic reflections are generated by static magnetic moments, whereas
magnetization is determined  
both by static magnetic moments and paramagnetic components. 
Therefore, the inconsistency appears. 
In Fig. 3, diffuse scattering is apparent at $2 \theta = 20 \sim 34^{\circ}$ 
in low magnetic fields, 
but it disappears in $H \geq 1$ T.
The diffuse scattering stems from short-range magnetic correlations 
in the low-dimensional spin system.
With increasing magnetic field, 
the short-range magnetic correlations are weakened and 
are replaced by long-range magnetic correlations 
(the polarized paramagnetic state).

\begin{figure}
\begin{center}
\includegraphics[width=8cm]{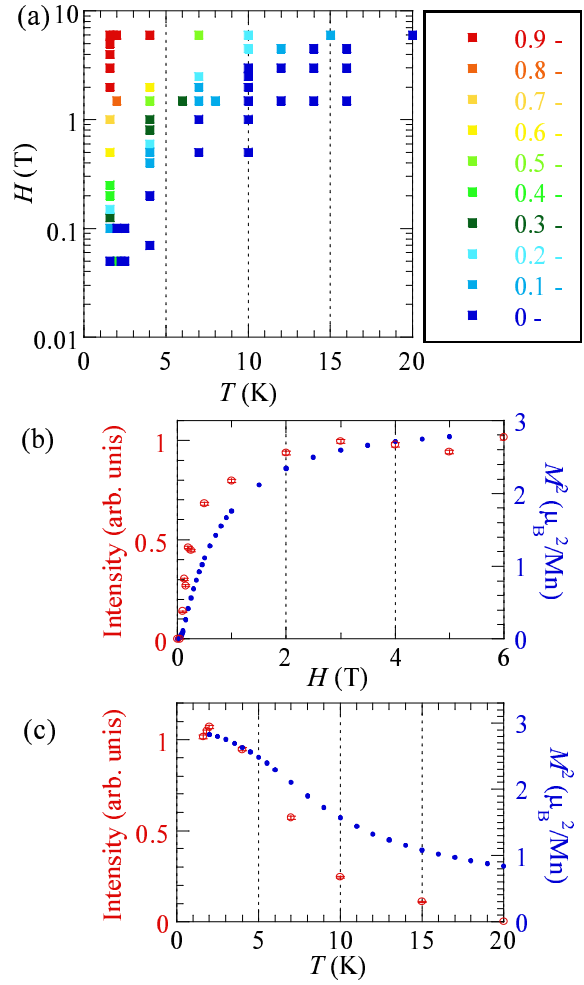}
\caption{
(Color online)
(a)
The integrated intensity map of the (0 1 1) reflection 
in the $T-H$ plane. 
The right panel shows intensity in arbitrary units. 
(b)
The magnetic-field dependence of  the integrated intensity 
of the (0 1 1) reflection at 1.6 K (open red circle). 
Closed blue circles show the square of magnetization measured at 1.8 K. 
(c)
Temperature dependence of  the integrated intensity 
of the (0 1 1) reflection in 6 T (open red circle). 
Closed blue circles represent square of magnetization measured in 5 T. 
}
\end{center}
\end{figure}

We obtained integrated intensities of nuclear reflections 
in the diffraction pattern at 1.6 K in 0 T. 
We calculated integrated intensities of 
nuclear reflections [$I_{\rm N}({\bf Q})$] 
using the following. 
\begin{equation}
I_{\rm N}({\bf Q}) = A |F_{\rm N}({\bf Q})|^2 n_{\bf Q}
\frac{1}{\sin 2 \theta \sin \theta}.
\end{equation}
Therein,  
\begin{equation}
F_{\rm N}({\bf Q}) = \sum_j b_j \exp (i {\bf Q} \cdot {\bf r}_j)
\exp (-B_j \frac{\sin^2 \theta}{\lambda^2}).
\end{equation}
$A$ is a scaling coefficient. 
$n_{\bf Q}$ is the number of reflections 
having the same intensity and $|{\bf Q}|$. 
$b_j$ is a scattering length of the $j$-th site atom. 
The values of $b_j$ are 
$7.02 \times 10^{-13}$,  $-3.75 \times 10^{-13}$, 
$5.13 \times 10^{-13}$, and $5.81 \times 10^{-13}$ cm for 
Sr, Mn, P, and O, respectively.\cite{ScatteringLength}  
${\bf r}_j$ is the position of the $j$-th site atom. 
$B_j$ is an atomic displacement parameter.  
We used atomic positions and 
values of $B_j$ determined in Rietveld refinements for 
a neutron diffraction pattern at 4.0 K in 0 T.\cite{Hase11a}  
Figure 5(a) portrays integrated intensities of nuclear reflections 
in the diffraction pattern at 1.6 K in 0 T 
versus 
corresponding calculated integrated intensities of nuclear reflections 
at $2 \theta < 55^{\circ}$. 
We obtained $A = 2.25 \times 10^{24}$ cm$^{-2}$. 

We obtained integrated intensities of reflections 
in the diffraction pattern at 1.6 K in 6 T.
The intensities include nuclear and magnetic contributions.  
We defined that   
the difference of intensities between 6 T and 0 T at each reflection
was an integrated intensity of a magnetic reflection 
in magnetization-plateau fields. 
We calculated the integrated intensities of 
magnetic reflections [$I_{\rm M}({\bf Q})$] 
in a neutron powder diffraction pattern 
using the following: 
\begin{equation}
I_{\rm M}({\bf Q}) = A |F_{\rm M}({\bf Q})|^2 n_{\bf Q}
\frac{1}{\sin 2 \theta \sin \theta},
\end{equation}
where 
\begin{equation}
F_{\rm M}({\bf Q}) = - \frac{g}{2} \frac{\gamma e^2}{m_e c^2}
\sum_j f(Q)_j 
\langle S_{jz} \rangle
\exp (i {\bf Q} \cdot {\bf r}_j)
\exp (-B_j \frac{\sin^2 \theta}{\lambda^2}).
\end{equation}
The $g$ value is 1.98. 
The value of $\frac{\gamma e^2}{m_e c^2}$ is 
$5.39 \times 10^{-13}$ cm. 
$f(Q)_j$ is a magnetic form factor of a $j$-th site ion.\cite{FormFactor} 
The direction of magnetic fields 
is always almost perpendicular to ${\bf Q}$ in the experimental setup. 
As described, in a plateau ground state, 
a magnetic moment on each site is fully polarized 
parallel and anti-parallel to external magnetic fields.  
Therefore, magnetic moments are 
always almost perpendicular to ${\bf Q}$. 
In Eq. (4), we can input the scalar $\langle S_{jz} \rangle$ 
instead of the vector ${\bf S}_{j \perp {\bf Q}}$ 
(components of ${\bf S}$ perpendicular to ${\bf Q}$).


We assume that 
the plateau ground state in SrMn$_3$P$_4$O$_{14}$
is close to that in the isolated AF spin-5/2 trimer 
as shown by the ellipse in Fig. 1. 
As described previously, 
in the plateau ground state in the isolated trimer, 
$\langle S_{1z} \rangle = -25/14$ 
on Mn1 site and $\langle S_{2z} \rangle = 15/7$ on Mn2 site, 
satisfying  
$ \langle S^{\rm T}_z \rangle = 
\langle S_{1z} \rangle + 2 \langle S_{2z} \rangle = 5/2$.
In a classical perspective, where magnitude of spin is constant, 
$ \langle S_{1z} \rangle = -5/2$ and 
$ \langle S_{2z} \rangle = 5/2$, 
also satisfying $ \langle S^{\rm T}_z \rangle = 5/2$.
We can calculate integrated intensities of magnetic reflections  
without any refined parameters 
except for assumptions of $\langle S_{jz} \rangle$ values. 
Figure 5(b) presents calculated results 
versus 
experimental integrated intensities of magnetic reflections 
at $2 \theta < 55^{\circ}$. 
The calculated results of the isolated trimer model 
agree well with the experimental results, whereas 
the calculated results of the classical model  
are larger than the experimental results. 
The value of $ \langle S^{\rm T}_z \rangle $ is constrained to be 5/2 
because of the 1/3 quantum magnetization plateau.  
In the constraint, 
the calculated intensity of each magnetic diffraction 
decreases monotonically with decreasing 
$\langle S_{2z} \rangle$  value. 
The set of $ \langle S_{1z} \rangle = -25/14$ and 
$ \langle S_{2z} \rangle = 15/7$ 
is a unique solution that can explain the experimental results 
within experimental errors.
Agreement between experimental and calculated results 
indicate that 
the plateau ground state in SrMn$_3$P$_4$O$_{14}$
is close to that in the isolated AF spin-5/2 trimer 
and that
the ground state is almost unchanged by 
the other exchange interactions, except for the $J_1$ interaction. 

We succeeded in direct observation of the plateau ground state  
in SrMn$_3$P$_4$O$_{14}$. 
Direct observation of the ground state 
was also conducted in Ca$_3$CuNi$_2$(PO$_4$)$_4$ in 
0 T.\cite{Pomjakushin07,Pomjakushin14,Podlesnyak07} 
The spin system is weakly coupled Ni-Cu-Ni trimers. 
In an isolated Ni-Cu-Ni trimer with single ion anisotropy 
$d (S_{{\rm Ni} z})^2 ~(d < 0)$, 
the ground states in the zero field are a doublet with 
$ \langle S^{\rm T}_z \rangle = \pm 3/2$. 
In the realistic model including effects of 
molecular fields from neighboring trimers, 
the unique expectation values of spin were calculated as 
$ \langle S_{{\rm Ni} z} \rangle = 0.9$ and 
$ \langle S_{{\rm Cu} z} \rangle = -0.3$.  
Ca$_3$CuNi$_2$(PO$_4$)$_2$ shows an AF order 
with an unusual multi-$k$ magnetic structure below $T_{\rm N} = 20$ K. 
The calculated expectation values of spin  
are in 10 \% accuracy in agreement with experimental values of 
ordered spins. 

In SrMn$_3$P$_4$O$_{14}$, 
the magnitudes of the ordered moments are
3.44 and 3.56 $\mu_{\rm B}$ on Mn1 and Mn2 sites, respectively, 
at 1.5 K in 0 T.\cite{Hase11a} 
The ordered moments on Mn1 and Mn2 sites are 
almost mutually anti-parallel. 
The magnitude of the Mn1 moment is close to 
$g \mu_{\rm B} | \langle S_{1z} \rangle | = 3.54 \mu_{\rm B}$, whereas 
the magnitude of the Mn2 moment is smaller than 
$g \mu_{\rm B} | \langle S_{2z} \rangle | = 4.24 \mu_{\rm B}$. 
The sublattice magnetizations are not fully saturated at 1.5 K.\cite{Hase11a} 
The ground states in the isolated Mn2-Mn1-Mn2 trimer 
is six-folded in 0 T.  
Therefore, the ground states with 
$ \langle S^{\rm T}_z \rangle = \pm 3/2$ or 
$ \langle S^{\rm T}_z \rangle = \pm 1/2$ 
might affect the experimental values of the magnetic moments. 

\begin{figure}
\begin{center}
\includegraphics[width=6cm]{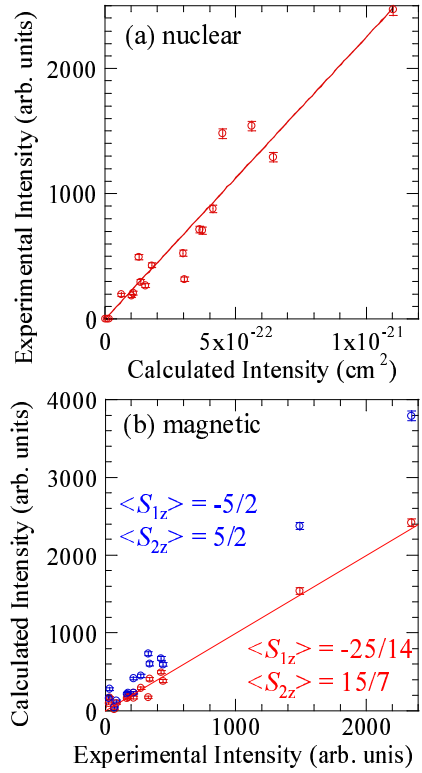}
\caption{
(Color online)
(a)
Integrated intensities of nuclear reflections
in the diffraction pattern at 1.6 K in 0 T versus 
calculated integrated intensities of nuclear reflections 
at $2 \theta < 55^{\circ}$. 
(b) 
Integrated intensities of magnetic reflections 
calculated for 
$ \langle S_{1z} \rangle = -25/14$ and 
$ \langle S_{2z} \rangle = 15/7$ (red) and 
$ \langle S_{1z} \rangle = -5/2$ and 
$ \langle S_{2z} \rangle = 5/2$ (blue)  
versus 
experimental integrated intensities of magnetic reflections 
at $2 \theta < 55^{\circ}$. 
}
\end{center}
\end{figure}

\section{Conclusion}

We performed neutron diffraction measurements 
on SrMn$_3$P$_4$O$_{14}$ pellets in magnetic fields 
to investigate the ground state in magnetization-plateau fields directly. 
The spin system is weakly coupled spin-5/2 trimers.  
Magnetic reflections indicating the occurrence of 
the coplanar spiral magnetic structure 
disappear in weak magnetic fields of $H \geq 0.3$ T. 
The result is consistent with the generation of  
the magnetic structure by weak exchange interactions 
(4.0 K at most) in the quasi-one-dimensional spin system. 
New magnetic reflections with integer indices appear in $H \geq 0.05$ T. 
They are correlated with the 1/3 quantum magnetization plateau. 
The temperature dependence of the integrated intensity of 
the new magnetic reflections
differs from that of an order parameter. 
Therefore, the state in $H \geq 0.3$ T is a polarized paramagnetic state 
without spontaneous magnetic order. 
The integrated intensities of the magnetic reflections 
calculated using the expectation values of spins in the plateau ground state 
in the isolated-trimer model 
agree well with
those obtained experimentally 
in the magnetization-plateau fields.  
We succeeded in direct observation of the plateau ground state  
in SrMn$_3$P$_4$O$_{14}$. 

\begin{acknowledgment}

This work was supported by KAKENHI (No. 23540396) and 
by grants from NIMS.
The neutron powder diffraction experiments were conducted
at SINQ, PSI Villigen, Switzerland
(Proposal Nos. 20111258 and 20130552). 
We are grateful 
to K. Kaneko for invaluable discussion and 
to S. Matsumoto for producing pellets. 

\end{acknowledgment}


\begin{thebibliography}{9}
\bibitem{Comment1}
Strictly speaking, we cannot see a ground state itself at finite temperatures 
because of thermal mixing of excited states. 
However, the influence of excited states is negligible 
in magnetization-plateau fields 
because of an energy gap.
\bibitem{Comment2}
As described herein, it is defined that 
the sign of a magnetic moment is positive (negative) 
when 
the sign of a spin value is positive (negative).
\end{thebibliography}
\end{document}